\documentstyle[epsfig,epsf]{article}
\newcommand{\ndt}{\noindent}

\def\nmnt {\nu_\mu \longleftrightarrow \nu_\tau}

\def\nm {\nu_\mu}

\def\Dm2{\Delta m^2}
\def\s2t{\sin^2 2\theta}

\def\be{\begin {equation}}
\def\ee{\end {equation}}

\begin{document}

\begin{center}
{\Large {\bf Summary of MACRO results on exotic physics}}
\end{center}

\vskip .7 cm

\begin{center}
Miriam Giorgini \par~\par
{\it Dept of Physics, Univ. of Bologna and INFN, \\
V.le C. Berti Pichat 6/2, Bologna, I-40127, Italy\\} 

E-mail: miriam.giorgini@bo.infn.it

\par~\par
Invited talk at Workshop on Exotic Physics with Neutrino Telescopes, \\ 
Uppsala, Sweden, September 2006.

\vskip .7 cm
{\large \bf Abstract}\par
\end{center}

{\small
MACRO was a multi-purpose experiment that took data from 1989 to 2000, at 
the underground Laboratory of Gran Sasso (Italy). MACRO gave important 
results/limits concerning: (i) the oscillation of atmospheric 
neutrinos, also in the non-conventional scenario of violations of 
Lorentz invariance, (ii) the searches for exotic particles (supermassive 
GUT magnetic monopoles, nuclearites, WIMPs), (iii) muon physics and 
astrophysics. A summary of the MACRO results will be presented and 
discussed, focusing the attention on the exotica searches.}

\vspace{5mm}

\section{Introduction}\label{sec:intro}
MACRO was a large area multipurpose underground detector \cite{r1} 
designed to search for rare events and rare phenomena in the cosmic radiation.
The detector was located in Hall B of the underground Gran Sasso Laboratory 
(Italy). It was optimised to search for the supermassive magnetic monopoles 
\cite{mono,pdecay} predicted by Grand Unified Theories (GUT).
The experiment obtained important results on atmospheric neutrino oscillations 
\cite{high,low,sterile,scatt,ultimo} and performed neutrino astronomy studies 
\cite{nuastro}, indirect searches for WIMPs \cite{wimps},
 search for low energy stellar gravitational collapse neutrinos \cite{grcol}, 
studies of the high energy underground muon flux (which is an indirect tool 
to study the primary cosmic ray composition and high energy hadronic 
interactions \cite{cr}), searches for fractionally charged particles (LIPs) 
\cite{lips} and other rare particles that may exist in the cosmic radiation. 

The detector started data taking in 1989 and it was running until December 
2000. The apparatus had global dimensions of $76.6 \times 12 \times 9.3$ 
m$^3$ and was composed of three sub-detectors: liquid scintillation counters, 
limited streamer tubes and nuclear track detectors. Each one of them could 
be used in ``stand-alone'' and in ``combined'' mode.
It may be worth to stress that all the physics and
astrophysics items listed in the 1984 proposal were covered and good results 
were obtained on each of them.

\section{Atmospheric neutrino oscillations}
\label{sec:nu-osc}

MACRO detected $\nm$-induced muon events in 4 different topologies. 

The {\it upthroughgoing muons} come from $\nm$ interactions in the rock below 
the detector; the $\nm$'s have a median energy of $\sim 50$ GeV.

\begin{figure*}[t]
\begin{minipage}[t]{0.4\linewidth}
\hspace{-1cm}
\epsfig{file=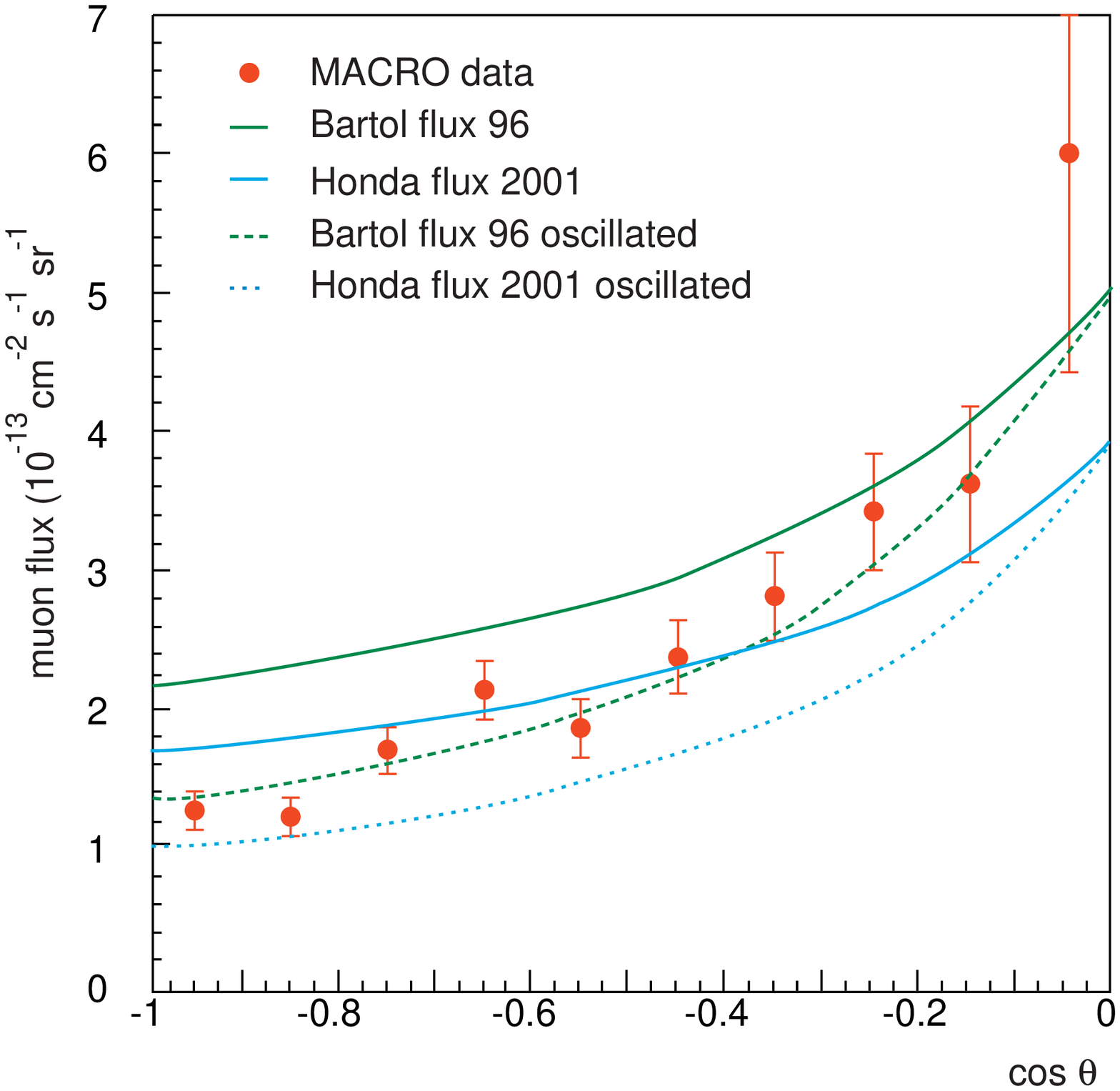,width=7.3cm,height=6cm}
\caption{Comparison of the MACRO upward-throughgoing muons (black circles) 
with the predictions of the Bartol96 and of the Honda2001 MC oscillated 
and non oscillated fluxes (oscillation parameters $\Dm2 = 2.3 \cdot 10^{-3}$ 
eV$^2$ and $\s2t_m =1$).}
\label{fig:zenith}
\end{minipage}\hfill
\begin{minipage}[t]{0.55\linewidth}
\centering\epsfig{file=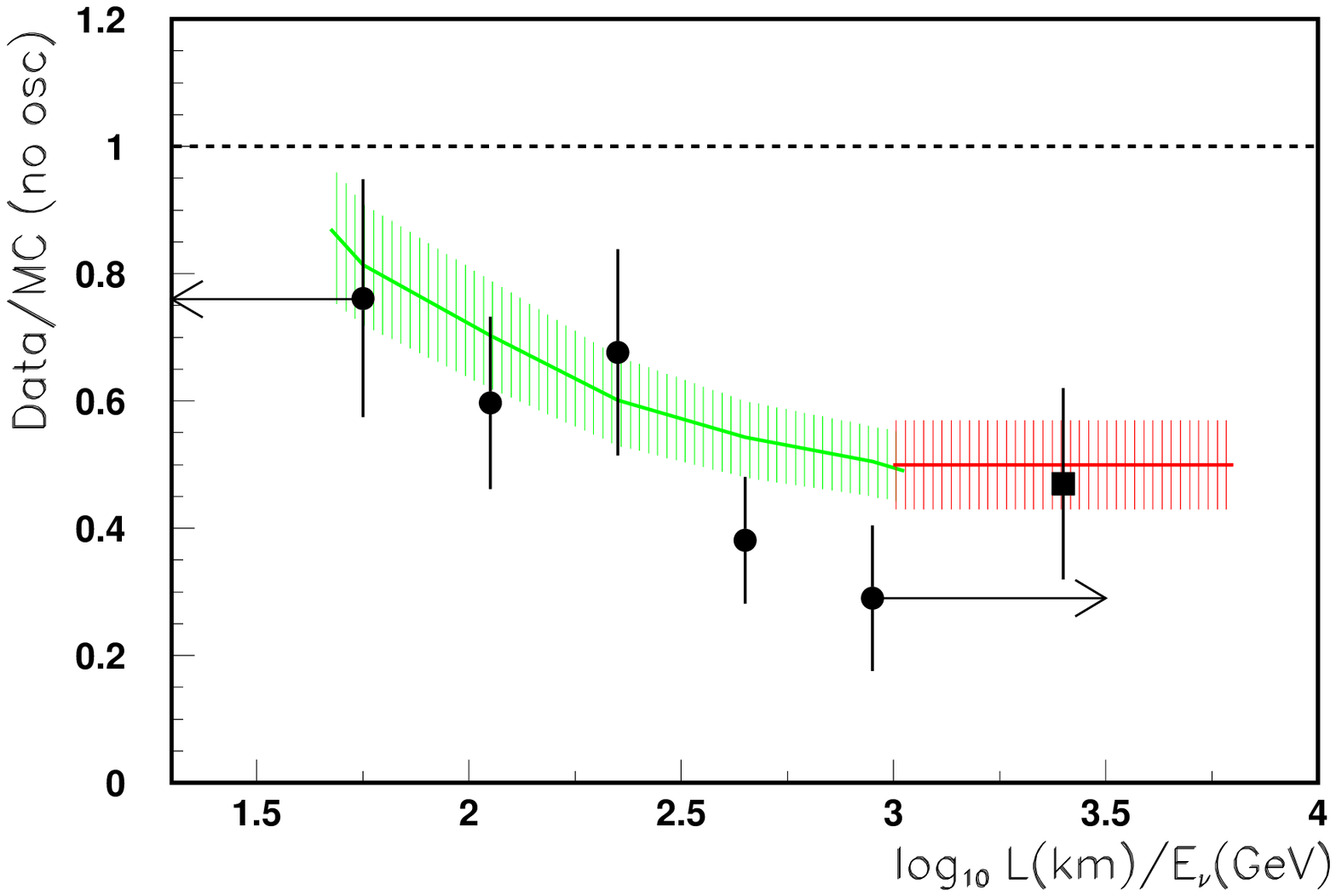,width=\linewidth}
\caption{Ratio Data/MC$_{\mbox{no~osc}}$ 
as a function of the estimated $L/ E_\nu$ for the upthroughgoing muon 
sample (black points). The solid line is the MC expectation assuming 
$\Dm2 = 2.3 \cdot 10^{-3}$ eV$^2$ and $\s2t_m = 1$. The last point 
(black square) is obtained from the IU sample.}
\label{fig:le}
\end{minipage}
\end{figure*}

Fig. \ref{fig:zenith} shows the zenith distribution of the measured 
902 upthroughgoing muons (black circles) compared with two 
MonteCarlo (MC) predictions: the Bartol96 \cite{bartol} flux with and 
without oscillations (the dashed and solid lines, respectively) and the 
Honda2001 flux \cite{honda}. The FLUKA MC predictions \cite{fluka} agree 
perfectly with the Honda2001.  

For a subsample of $\sim 300$ upthroughgoing events, we estimated the muon 
energy through Multiple Coulomb Scattering in the rock absorbers in the
lower apparatus \cite{scatt}. The evaluated resolution on $E_\nu$ is 
$\sim 100 \%$. The parent neutrino path length is $L \sim 2 R_E \cos \Theta$, 
where $R_E$ is the Earth radius. Fig. \ref{fig:le} shows the 
ratio data/MC as a function of the estimated $L/E_\nu$ for the 
upthroughgoing muon sample. 
The black circles are data/Bartol96 MC (assuming no oscillations); the
solid line is the oscillated MC prediction for $\Dm2 = 2.3 \cdot 10^{-3}$ 
eV$^2$ and $\s2t_m =1$. The shaded region represents the simulation 
uncertainties. The last point (black square) is obtained
from semicontained upward going muons.

The low energy events ($IU$, $ID+UGS$ \cite{low}) are produced 
by parent $\nm$ interacting inside the lower detector, or by upgoing muons 
stopping in the detector. The median energy of the parent neutrino is 
$\sim 3-4$ GeV for all topologies. In both cases, the zenith distributions are 
in agreement with the oscillation prediction with the optimised parameters
\cite{ultimo}. 

In order to reduce the effects of systematic uncertainties in the MC 
absolute fluxes we used the following three independent ratios 
\cite{ultimo}:
\begin{enumerate}
        \item [(i)] High Energy data: vertical/horizontal ratio,
        $R_{1} = N_{vert}/N_{hor}$
        \item [(ii)] High Energy data: low energy/high energy ratio,
        $R_{2} = N_{low}/N_{high}$
        \item [(iii)] Low Energy data: 
        $R_{3} = (Data/MC)_{IU}/(Data/MC)_{ID+UGS}$
\end{enumerate}
Combining the three independent results, the no oscillation hypothesis 
is ruled out at the $\sim 5 \sigma$ level ($6 \sigma$ if the absolute 
values compared to the Bartol96 flux are used) \cite{trieste}. 

To evaluate the hypothesis of oscillation for different values of $\Dm2$ and 
 $\s2t_m$, the Feldman-Cousins \cite{fel_cou} procedure was used and the 
corresponding $90\%$ C.L. region for the $\nmnt$ oscillation is 
given in ref \cite{ultimo}. The best fit is reached at $\Dm2 = 
2.3 \cdot 10^{-3}$ eV$^2$ and $\s2t_m = 1$.

\subsection{Search for exotic contributions to atmospheric neutrino 
oscillations} \label{sec:lorentz}

MACRO searched for ``exotic'' contributions to standard mass-induced
atmospheric neutrino oscillations, arising from a possible violation of Lorentz
invariance (VLI) , using two different and complementary analyses. The first 
approach uses the low energy ($E_\nu < 28$ GeV) and the high energy 
($E_\nu > 142$ GeV) samples. The mass neutrino
oscillation parameters have the values given in Sect. \ref{sec:nu-osc} and 
 we mapped the evolution of the $\chi^2$ estimator
in the plane of the VLI parameters $\Delta v$ and $\sin^2 2 \theta_v$. No
$\chi^2$ improvement was found, so we applied the Feldman-Cousins 
\cite{fel_cou} method to determine $90\%$ C.L. limits on the parameter:
$|\Delta v| < 3 \cdot 10^{-25}$ \cite{lorentz}.

The second approach exploits a data subsample
characterised by intermediate neutrino energies. It is based on the maximum
likelihood technique and 
considers the mass neutrino oscillation parameters varying in the $90\%$ C.L. 
border \cite{ultimo}. The obtained $90\%$ C.L. limit 
on the $\Delta v$ parameter is also around $10^{-25}$ \cite{praga} .

\section{\label{sec:nu-astrophysical} Neutrinos from astrophysical sources}
\subsection{\label{sec:HE} Search for astrophysical HE muon neutrinos}  
High energy $\nm$'s are expected to come from several galactic 
and extra-galactic sources. An excess of events was searched for around 
the positions of known sources in $3^{\circ }$ (half width) angular bins. The 
$90\%$ C.L. upper limits on the muon fluxes from specific celestial sources
were in the range $10^{-15} \div 10^{-14}$ cm$^{-2}$ s$^{-1}$ 
\cite{mudiffu}. A search for 
time coincidences of the upgoing muons with $\gamma$-ray bursts was also 
made. No statistically significant time correlation was found \cite{nuastro}.

A different analysis was made for the search for a diffuse astrophysical 
neutrino flux, using a dedicated method to select higher energy upthroughgoing
muons. The flux upper limit was set at the level of $1.5 \cdot 10^{-14}$ 
cm$^{-2}$ s$^{-1}$ \cite{nudiffu}.

\subsection{\label{sec:WIMPs} Indirect searches for WIMPs}
Weakly Interacting Massive Particles (WIMPs) could be part of the
galactic dark matter; they could be intercepted by celestial bodies,
slowed down and trapped in their centres, where WIMPs and anti-WIMPs could
annihilate and yield neutrinos of GeV or TeV energy,
in small angular windows from their centres. One WIMP candidate 
is the lowest mass neutralino.

To look for a WIMP signal, we searched for upthroughgoing muons from the 
Earth centre, using $10^{\circ} \div 15^{\circ}$ cones around the Nadir; the 
$90\%$ C.L. muon flux limits are $0.8 \div 1.4 \cdot 10^{-14}$ 
cm$^{-2}$ s$^{-1}$ \cite{wimps}. These limits, when compared with 
the predictions of a supersymmetric model, eliminate a sizable range 
of parameters used in the model. 

A similar procedure was used to search for $\nm$ from the
Sun: the muon upper limits are at the level of about 
$1.5 \div 2 \cdot 10^{-14}$ cm$^{-2}$ s$^{-1}$ \cite{wimps}.

\subsection{\label{sec:GC} Neutrinos from stellar gravitational collapses} 
A stellar gravitational collapse of the core of a massive star is expected 
to produce a large burst of all types of neutrinos and antineutrinos with 
energies of $ 5 \div 60$ MeV and with a duration of $\sim 10$ s. No stellar 
gravitational collapses in our Galaxy were observed from 1989 to 2000 
\cite{grcol}. 

\begin{figure*}[t]
\begin{minipage}[t]{0.5\linewidth}
\hspace{-1cm}
\epsfig{file=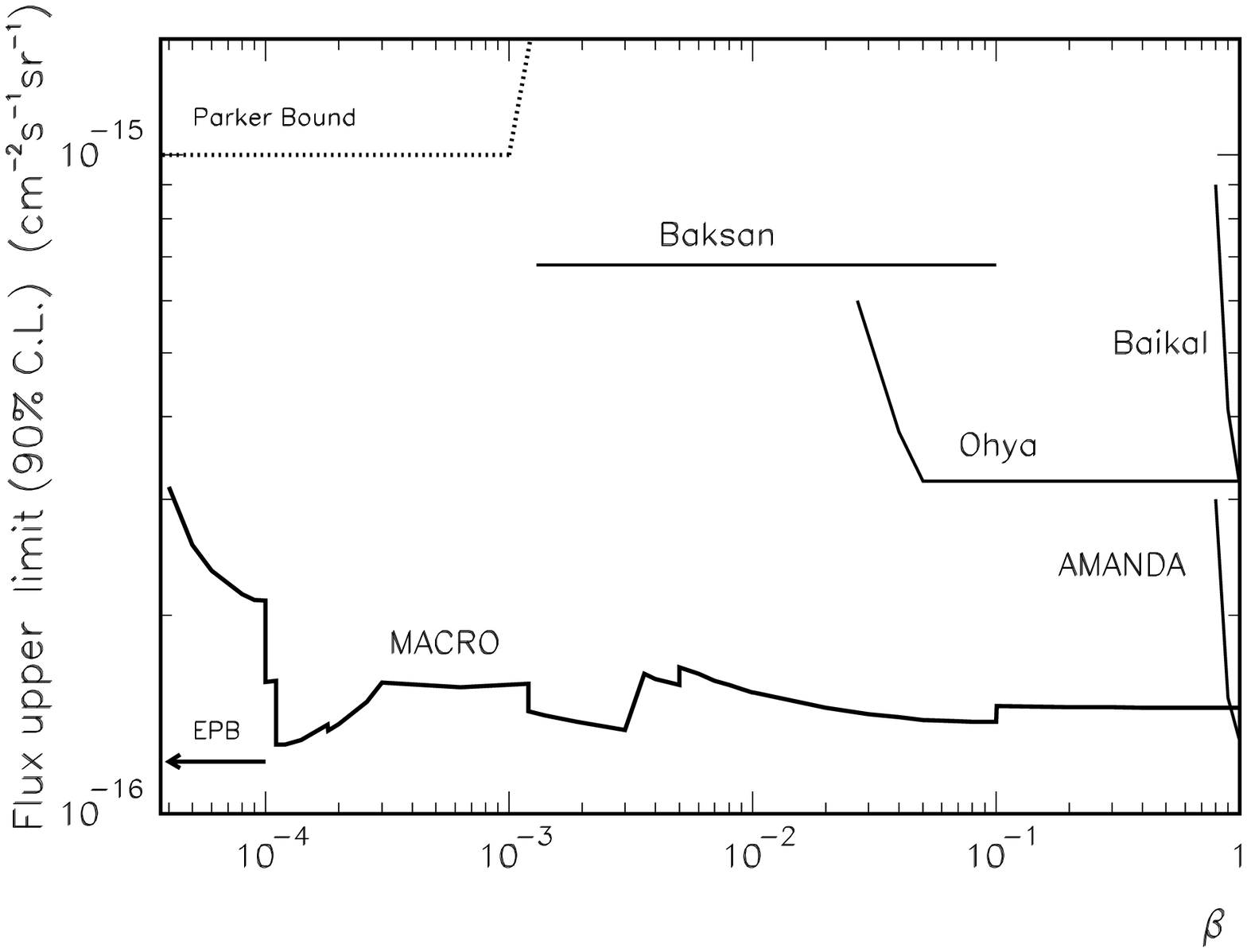,width=9cm,height=8cm}
\caption{$90\%$ C.L. upper limit obtained by MACRO for an isotropic flux 
of GUT MMs with $g=g_D$ compared with direct limits given by other 
experiments.}
\label{fig:mm}
\end{minipage}\hfill
\begin{minipage}[t]{0.47\linewidth}
\hspace{-0.4cm}
\centering\epsfig{file=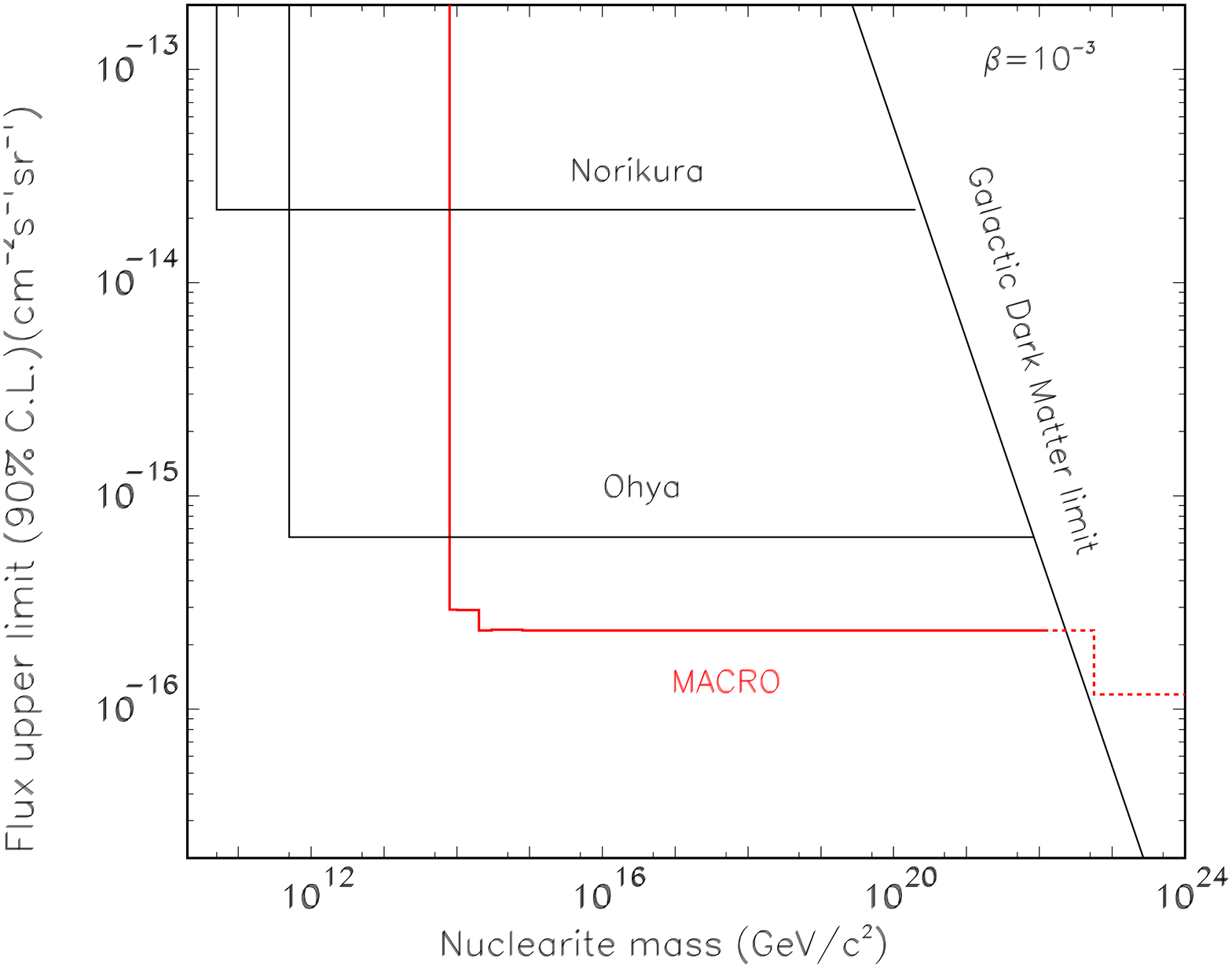,height=6.5cm,width=8cm}
\caption{$90\%$ C.L. upper limits versus mass for downgoing nuclearites
with $\beta=2 \cdot 10^{-3}$ at ground level. The MACRO limit for nuclearite 
masses larger than $5 \cdot 10^{22}$ GeV/c$^2$ has been extended and 
corresponds to an isotropic flux.}
\label{fig:nucl}
\end{minipage}
\end{figure*}

\section{\label{sec:exotic} Search for exotic particles}
\subsection{\label{sec:MM} Search for GUT magnetic monopoles (MMs)} 
Supermassive magnetic monopoles predicted by Grand Unified Theories
(GUT) of the electroweak and strong interactions should have 
masses $m_M \sim 10^{17}$ GeV. 

MACRO was optimised to search for an isotropic flux of GUT MMs in the cosmic
radiation. The three sub-detectors had sensitivities in different $\beta$ 
regions, covering the velocity range $4 \cdot 10^{-5} < \beta < 1$.
They allowed multiple signatures of the same rare event candidate. No 
candidates were found by any of the three subdetectors. Fig. \ref{fig:mm} 
shows the $90\%$ C.L. flux upper limits for $g=g_D$ poles (one unit of Dirac 
magnetic charge) plotted versus $\beta$ \cite{mono} together with direct 
limits set by other experiments \cite{mono_altri}. The MACRO MM direct limits
 are by far the best existing over a very wide range of $\beta$.

The interaction of the GUT monopole core with a nucleon can lead to a 
reaction in which the nucleon decays, $M+p \rightarrow M + e^+ + \pi^0$. 
 MACRO dedicated an analysis procedure to detect nucleon decays induced 
by the passage of a GUT MM in the streamer tube system (a fast $e^+$
track from a slow ($\beta \sim 10^{-3}$) MM track). The $90\%$ C.L. flux 
upper limits established by this search are at the level of 
$\sim 3 \cdot 10^{-16}$ cm$^{-2}$ s$^{-1}$ sr$^{-1}$ for 
$10^{-4} \leq \beta \leq 0.5 \cdot 10^{-2}$; they are valid for catalysis 
cross sections $5\cdot 10^2 < \sigma_{cat} < 10^3$ mb \cite{pdecay}. 

\subsection{\label{sec:nucl} Search for nuclearites, Q-balls and LIPs} 
Strangelets should consist of aggregates of $u,~d$ and $s$ quarks in almost 
equal proportion \cite{nucl} and would have typical galactic velocities 
$\beta \sim 10^{-3}$. The MACRO $90\%$ C.L. upper limits for an isotropic 
flux of nuclearites with $10^{-5} \le \beta \le 1$ was at the level of 
$1.5 \cdot 10^{-16}$ cm$^{-2}$ s$^{-1}$ sr$^{-1}$ \cite{nuclea}.

MACRO searched also for charged Q-balls (aggregates of squarks, sleptons 
and Higgs fields) \cite{qballs}, giving an upper limit of 
$\sim 10^{-16}$ cm$^{-2}$ s$^{-1}$ sr$^{-1}$ \cite{qballs2}.

Fractionally charged particles could be expected in Grand Unified 
Theories as deconfined quarks; the expected charges range 
from $Q=e/5$ to $Q= 2/3e$. LIPs should release a fraction $(Q/e)^2$ of the
energy deposited by a muon traversing a medium. The $90\%$ C.L. flux upper 
limits for LIPs with charges $e/3,~2/3e$ and $e/5$ are at the level 
of $10^{-15}$ cm$^{-2}$ s$^{-1}$ sr$^{-1}$ \cite{lips}.

\section{\label{sec:conclu} Conclusions}

$-$ Standard atmospheric $\nm$ oscillations: no-oscillation hypothesis 
ruled out at $5 \div 6 \sigma$. 

Best fit parameters: 
 $\Dm2 = 2.3 \cdot 10^{-3}$ eV$^2$ and $\s2t_m =1$. \par
\ndt $-$ VLI: $|\Delta_v|$ upper limits of the order of $10^{-25}$. \par
\ndt $-$ MM search: upper flux limit of $1.4 \cdot 10^{-16}$ 
cm$^{-2}$ s$^{-1}$ sr$^{-1}$ for $4 \cdot 10^{-5} < \beta < 1$.  \par
\ndt $-$ Nuclearite search: upper flux limit of $10^{-16}$  
cm$^{-2}$ s$^{-1}$ sr$^{-1}$ for $\beta \simeq 10^{-3}$. \par
\ndt $-$ Charged Q-balls search: upper flux limit of $\sim 10^{-16}$ 
cm$^{-2}$ s$^{-1}$ sr$^{-1}$. \par
\ndt $-$ WIMP search: upper flux limit of $\sim 10^{-14}$ cm$^{-2}$ s$^{-1}$ 
sr$^{-1}$. \par 
\ndt $-$ LIP search: upper flux limit of $6.1 \cdot 10^{-16}$ 
cm$^{-2}$ s$^{-1}$ sr$^{-1}$. \par~\par

{\bf Acknowledgements} 

I acknowledge the cooperation 
of the members of the MACRO Collaboration and, in particular, of the 
Bologna group.

\end{document}